# Mean Estimate Distances for Galaxies with Multiple Estimates in NED-D

Ian Steer

176 The Esplanade, Ste. 705, Toronto, ON M5A 4H2, Canada; steer@bell.net

## Abstract

Numerous research topics rely on an improved cosmic distance scale (e.g., cosmology, gravitational waves), and the NASA/IPAC Extragalactic Database of Distances (NED-D) supports those efforts by tabulating multiple redshift-independent distances for 12,000 galaxies (e.g., Large Magellanic Cloud (LMC) zero-point). Six methods for securing a mean estimate distance (MED) from the data are presented (e.g., indicator and Decision Tree). All six MEDs yield surprisingly consistent distances for the cases examined, including for the key benchmark LMC and M106 galaxies. The results underscore the utility of the NED-D MEDs in bolstering the cosmic distance scale and facilitating the identification of systematic trends.

*Unified Astronomy Thesaurus concepts:* Cosmology (343); Galaxy distances (590); Distance indicators (394); Galaxies (573); Extragalactic astronomy (506);

*Supporting material:* machine-readable tables

## 1. Introduction

Redshift-independent distances (hereafter distances) tied to multiple indicators are beneficial for gravitational wave cross-matching, and establishing the cosmic distance scale, peculiar velocity flows, and the Hubble constant. Consequently, the NASA/IPAC Extragalactic Database of Distances (NED-D) was created in part to serve as a resource that hosts such pertinent information (Steer et al. 2017). NED-D is the largest compilation of extragalactic distances, containing the majority of published estimates since 1980. Currently, distances for 150,000 galaxies are available, and 12,000 of those have multiple distances based on a total 78,000 estimates. Those estimates are tied to at least 77 separate indicators.

Mean distances presently cited in NED are inferred from an unweighted average of all distances per galaxy, as published. No corrections are applied to account for differences in zero-point or distance indicators, nor are outliers removed. The key objective of the present study is to report on the implementation of diverse methods for estimating the mean distance, thereby establishing a multifaceted mean estimate distance (MED) procedure. For example, a best estimate distance (BED, Harris et al. 2010) approach applied to Cen A (NGC 5128) resulted in an error-weighted mean of 3.8 ± 0.3 Mpc. Specifically, Harris et al. 2010 selected the single most precise and recent distance for each of four selected primary indicators. Eight primary indicators and numerous additional distances are available for that galaxy in NED-D. Mean distances cited in certain other compilations also follow a BED approach, and for example, a weighted mean for three estimates based on two primary indicators is provided for M101 (NGC 5457) in the Extragalactic Distance Database (EDD; Tully et al. 2016). For that galaxy NED-D features 112 estimates based on six primary indicators. Other compilations include the Updated Nearby Galaxies Catalog (UCNG; Karachentsev, Makorov, & Kaisina 2013), and the HyperLEDA catalog (Makarov, Prugniel, & Terekhova 2014).

It is desirable that an enhanced MED approach be relayed to researchers when selecting extragalactic distance estimates. Six means shall be presented to researchers: an unweighted mean of the (1) distance estimates or (2) indicators; a weighted mean of the indicators based on either (3) distance error or (4) date of publication; and a Decision Tree mean involving either (5) a BED approach based on selected estimates per indicator or (6) a mean of indicators weighted by preference. A seventh mean that combines a subset of the aforementioned is likewise provided.

This study is organized as follows. In section 2, descriptions are provided for the different indicators and indicator categories, the placement of estimates onto a common scale, and the clipping of estimate outliers. The different MEDs are described in Section 3, and MEDs are evaluated for the LMC, the primary extragalactic distance scale zero-point, and to 40 Messier galaxies including M106, an alternate distance scale zero-point galaxy. Conclusions regarding the determination of mean distances for galaxies with multiple estimates are summarized in Section 4.

## 2. Distance Indicators, Distance Scales, and Outliers

The NED compilation of distances (NED-D) is described in Steer et al. 2017. Briefly, the distances include published peer-reviewed estimates since 1980, as well as some vetted non-peer-reviewed distances. At least 77 distance indicators are currently in use (Table 1), and hence NED facilitates the identification of systematic uncertainties.

The placement of indicators into categories has been revised. Distances were previously classified as primary standard candle, primary standard ruler, or secondary. A new indicator category is conveyed to recognize that certain indicators are neither primary nor secondary. The added category accounts for 19 tertiary indicators that are imprecise, and includes distances based on the Infrared Astronomical Satellite (IRAS) indicator. The precision of primary and secondary estimates is typically 10% and 20%, respectively. Accordingly, primary distances are weighted four times greater than secondary estimates (e.g., Tully et al. 2013).

Distances in NED can be tied to different scales, and can assume either a Hubble constant or LMC modulus. Distances can be placed onto a homogenized scale, however, by making use of the two ancillary data columns provided. Distances based on a Hubble constant offset from $H_0 = 70$ km s$^{-1}$ Mpc$^{-1}$ are noted in an ancillary column. Similarly, distances based on an LMC modulus offset from $\mu_0 = 18.50$ mag are likewise noted.

Published uncertainty estimates are inhomogeneous, and may be tied to a weighted approach, standard deviation, standard error, formal uncertainties tied to least-squares fitting routines, a quadrature sum of standard error and systematic uncertainties, etc. For this study the uncertainties are adopted verbatim and no adjustments are made for the aforementioned differences. For all MEDs computed the standard deviation is cited, even for weighted mean approaches.

### 3. Mean Distances for the LMC and M106

MEDs were determined once tertiary distances were discarded, primary and secondary distances were placed on a common scale, and 3$\sigma$ outliers were excluded. The MED methods employed are summarized in Table 2.

The unweighted mean of the 940 primary distance estimates to the LMC implies 49.57 ± 3.03 kpc (MED 1). The unweighted mean of the mean distances for each of 17 primary indicators is 50.08 ± 1.58 kpc (MED 2). The unweighted mean of the error-weighted mean distances for each of 17 indicators is 50.29 ± 1.37 kpc (MED 3), and importantly, the uncertainty cited here is the standard deviation rather than the canonical weighted uncertainty. The unweighted mean of the date-weighted mean distances for each of 17 indicators is 50.31 ± 1.46 kpc (MED 4). The date-weighting scheme adopted for MED 4 is $1.258^n$, where n is the number of years between publication and 1980. Estimated distances to the LMC, including means for each primary indicator and based on each MED, are shown in Figure 1.

Regarding method 4, attention has been drawn to the fact that over time, distances and indicators improve in precision and accuracy (e.g., Helou & Madore 1988, de Grijs, Wicker, & Bono 2014). Decade-over-decade improvement in the standard deviation among Hubble constants published since 1980 demonstrates this, as shown in Figure 2. Controversy in the 1980s over whether the Hubble constant was close to 50 or 100 km s$^{-1}$ Mpc$^{-1}$ has been reduced to whether it is closer to 68 or 73, depending on whether global values based on cosmological microwave background radiation (Bennett et al. 2013, Planck Collaboration et al. 2018), or local values based on Cepheids calibrated Type Ia supernovae are assumed (Freedman et al. 2012, Riess et al. 2016). Published Hubble constant estimates in the most recent decade favor the global value, but do not rule out the local one. Interestingly, one of the teams behind one of the local values has recently found more conclusive evidence for the global value, based on an independent calibration of the distance scale using the tip of the red giant (TRGB) indicator (Freedman et al. 2019).

To within the standard deviation, error weighting and date weighting do not impact the mean distance obtained. Two Decision Tree approaches were evaluated. The first follows a BED approach based on selecting the single most precise estimate per indicator, and in case of a tie the most recent. For the LMC, this provides a distance of 49.90 ± 1.78 kpc (MED 5). The second Decision Tree approach applies weighting each indicator based on a precomputed ranking. The subjective ranking is relayed in Table 1. For the LMC the result is 49.33 ± 1.48 kpc (MED 6). Both Decision Tree approaches thus also supply consistent distances to within the standard deviations.

Table 3 hosts the mean distances to the LMC based on each of the 17 indicators, and each MED. A seventh mean is likewise employed, and combines the literature preferred BED approach of selecting estimates (MED 5) with a weighted mean of MED 2, 3, and 4. Method 1 is not included because it exhibits the most scatter, and is the mean of distances regardless of indicators, while method 6 was excluded owing to the increased subjectivity. The unweighted (MED 2), error-weighted (MED 3), and date-weighted (MED 4) means were combined with weights of 1:2:4, respectively. The result is a combined MED 7 estimate of 50.09 ± 1.61 kpc (m-M = 18.499 ± 0.069 mag).

For the LMC all six methods and the combined method 7 produce a consistent distance, and the standard deviation for each MED is ~3%. The mean distances for the LMC support the Pietrzynski et al. (2019) finding. Pietrzynski et al. (2019) obtained 49.59 ± 0.09 (stat.) ± 0.54 (syst.) kpc based on 20 eclipsing binaries. The canonical LMC distance of 50.1 ± 2.5 kpc adopted by the Hubble Space Telescope Key Project (Freedman et al. 2001), with an accuracy of 5%, remains within 1% of the eclipsing binary determination as well as all but one of the MED estimates.

A comparison was likewise carried out on 40 Messier galaxies. M106 is of particular interest as an alternate zero-point, and the primary standard ruler megamaser-based distance is 7.54 ± 0.23 Mpc (Riess et al. 2016). For M106 there are 112 distances, of which 3 are tertiary and excluded. Another 3 are discarded as 3σ outliers, leaving 106 distances. Those include 8 distances based on megamasers, for which only the Riess et al. 2016 result is examined. The different MED methods again produce consistent distances to within the standard deviations, resulting in a mean for the six MEDs based on primary indicators of 7.49 ± 0.02 Mpc. Results are presented in Table 4, and displayed in Figure 3.

Table 5 hosts the mean distances to 40 Messier galaxies based on MEDs 2, 3, 4, and 5. In that table, primary and secondary distances are combined and weighted at 4:1. All mean distance estimates for the 40 galaxies were calculated manually. To determine MEDs for the entire ensemble of ~272,000 NED distances, a Python program was created by a visiting member of the NED Team (Michael Randall). The results were subsequently compared with redshift-based distances. The latter sample featured galaxies with distances greater than 5 Mpc, and heliocentric recessional velocities greater than 300 km s$^{-1}$. That excludes nearby galaxies with high peculiar velocities, and galaxies with low or negative recessional velocities. The comparison was also limited to galaxies with a heliocentric recessional velocity of v < 32,000 km s$^{-1}$, and with mean distances within 1,000 Mpc. Linear distances were determined assuming a Hubble constant of $H_0$ = 70 km s$^{-1}$ Mpc$^{-1}$. Note that only galaxies with multiple distance estimates are viable for MED evaluations.

For the 11,699 galaxies available, the mean redshift-based distance is 85.3 Mpc, and the mean redshift-independent distance from the six MEDs is 87.5 Mpc. Again, the six MEDs and the combined MED 7 provide distances consistent to within the standard deviations. Mean distances for the ensemble based on all six MEDs and an added seventh method are presented in the machine-readable version of Table 6, and are inferred from 78,228 eligible distances. A representative sample is shown here in Table 6, for guidance. A Hubble graph for the 11,699 galaxies, and their positions in galactic coordinates, are shown in Figures 4 and 5, respectively.

## 4. Summary and Discussion

Establishing reliable distances for galaxies with primary and secondary distances is a first step in calibrating indicators, providing an improved distance scale and Hubble constant, and aiding determination of the latter's evolution. As a result, the NED team evaluated six methods to estimate MEDs, with an aim in part to providing fellow researchers additional pertinent information that may facilitate the identification of systematic trends. Those MEDs are summarized in Table 2. All six MEDs produce consistent distances to within the standard deviations for the LMC, M106, for all 40 Messier galaxies, and in general among all galaxies with multiple distances in NED (n = 11,699).

In this first benchmarking of the cited approaches the MED distances determined for the LMC are consistent with one another and agree with the Pietrzynski et al. (2019) result to within 1%. For M106 the distances computed likewise are consistent and within 1% of the fiduciary megamaser-based estimate (Riess et al. 2016).

New distances will be provided for ~320,000 galaxies, and inferred from indicators which include the Fundamental Plane and Brightest Cluster Galaxy methods in an update planned for 120,000 galaxies with distances (Saulder et al. 2016). Those galaxies will benefit from MEDs, since each will possess on average four estimates based on at least two indicators.

Repeated consistency among MEDs in multiple applications increases confidence in the cosmological distance scale and the estimates it is based on. It indicates both are surprisingly free from unknown systematic errors, unless such unknowns cancel fortuitously. Overall, NED-D is pertinent for a diverse suite of research topics, such as aiding those on the Swope observatory team to quickly identify NGC 4993 as the host of gravitational wave GW170817 (Drout et al. 2017).

The anonymous referee was as important to this article as the author, who appreciates and admires the dedication. It has been an honor to serve with members of the NED Team, past and present, including Kay Baker, Ben H. P. Chan, Xi Chen, David Cook, Harold G. Corwin, Rick Ebert, Cren Frayer, George Helou, Jeff Jacobson, Joyce Kim, Tak Lo, Barry F. Madore, Joseph M. Mazzarella, Olga Pevunova, Michael Randall, Marion Schmitz, Scott Terek, Cindy Shin-Ywan Wang, and Xiuqin Wu. This research made much use of the NASA/IPAC Extragalactic Database (NED), which is operated by the Jet Propulsion Laboratory, California Institute of Technology, under contract with the National Aeronautics and Space Administration. Additional generous support to IS from the Carnegie Institution of Canada is also gratefully appreciated.

**ORCID iDs**

Ian Steer https://orcid.org/0000-0003-3716-858X

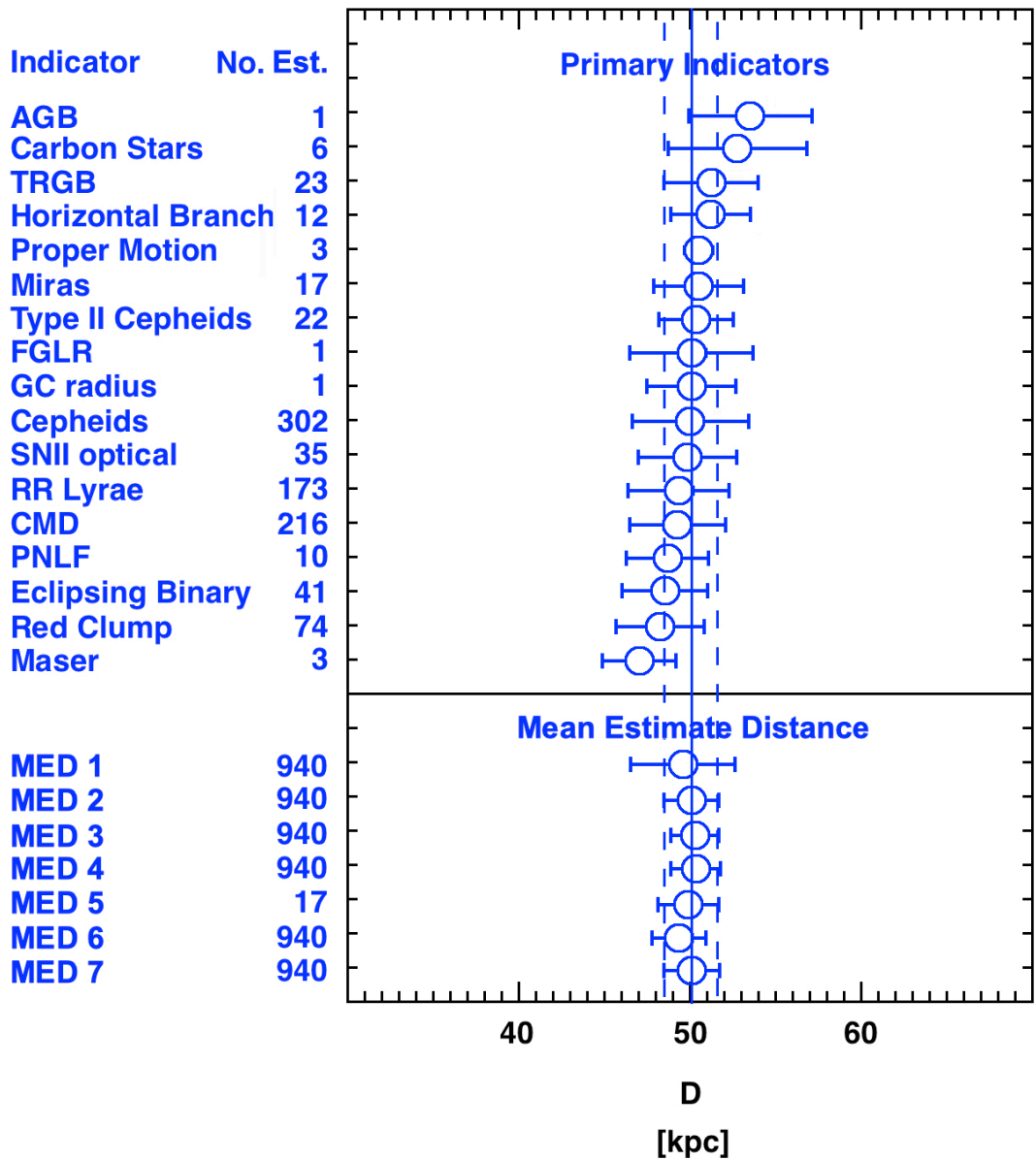

**Figure 1**. LMC mean distance (solid line) and 1σ standard deviation (dashed lines) from MED method 2, the unweighted mean of 17 primary indicators. Indicator means and standard deviations shown by data points and error bars, except indicators with one estimate such as AGB, where 1σ precision of individual estimate is shown. Mean estimate distances and standard deviations are shown for seven MED methods by lower data points and error bars.

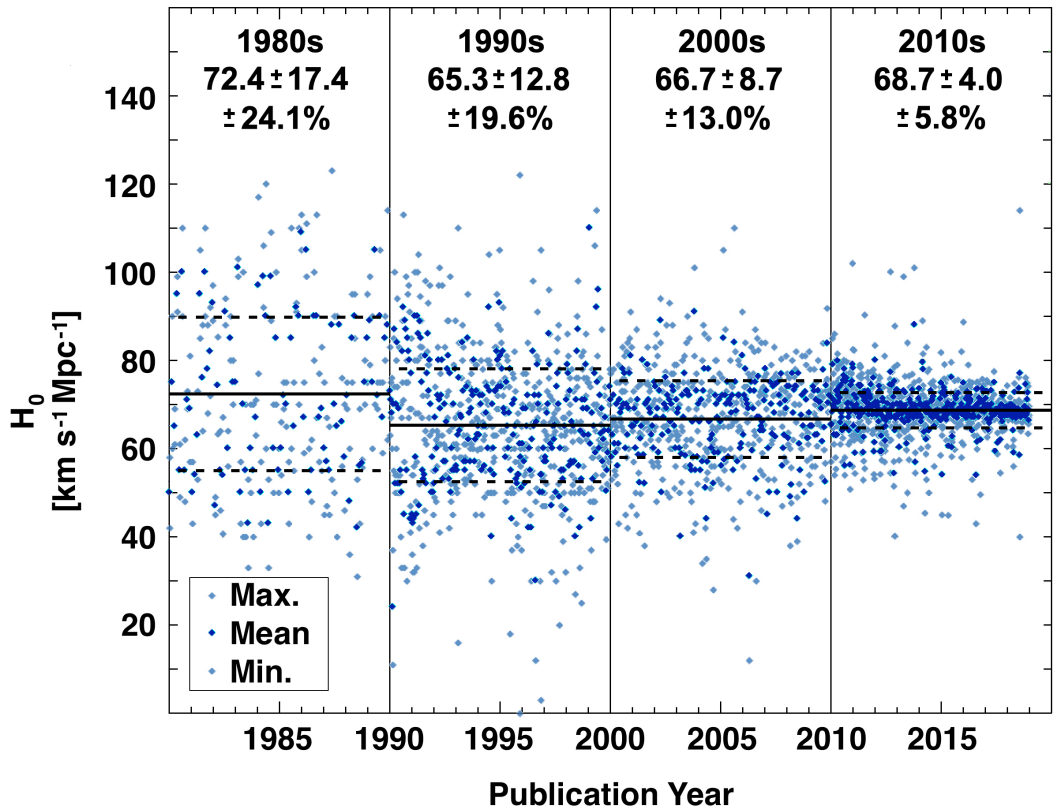

**Figure 2**. Improvement in the precision of extragalactic distances over time is evident in the decade-over-decade improvement in standard deviation among Hubble constant estimates published from 1980 to 2019 (n = 966). Standard deviation of individual estimates is shown where available by maximum and minimum values in light blue. Data are from an internally maintained update of the John Huchra Hubble constant database originally maintained for the NASA Hubble Space Telescope Key Project on the extragalactic distance scale (Freedman et al. 2001).

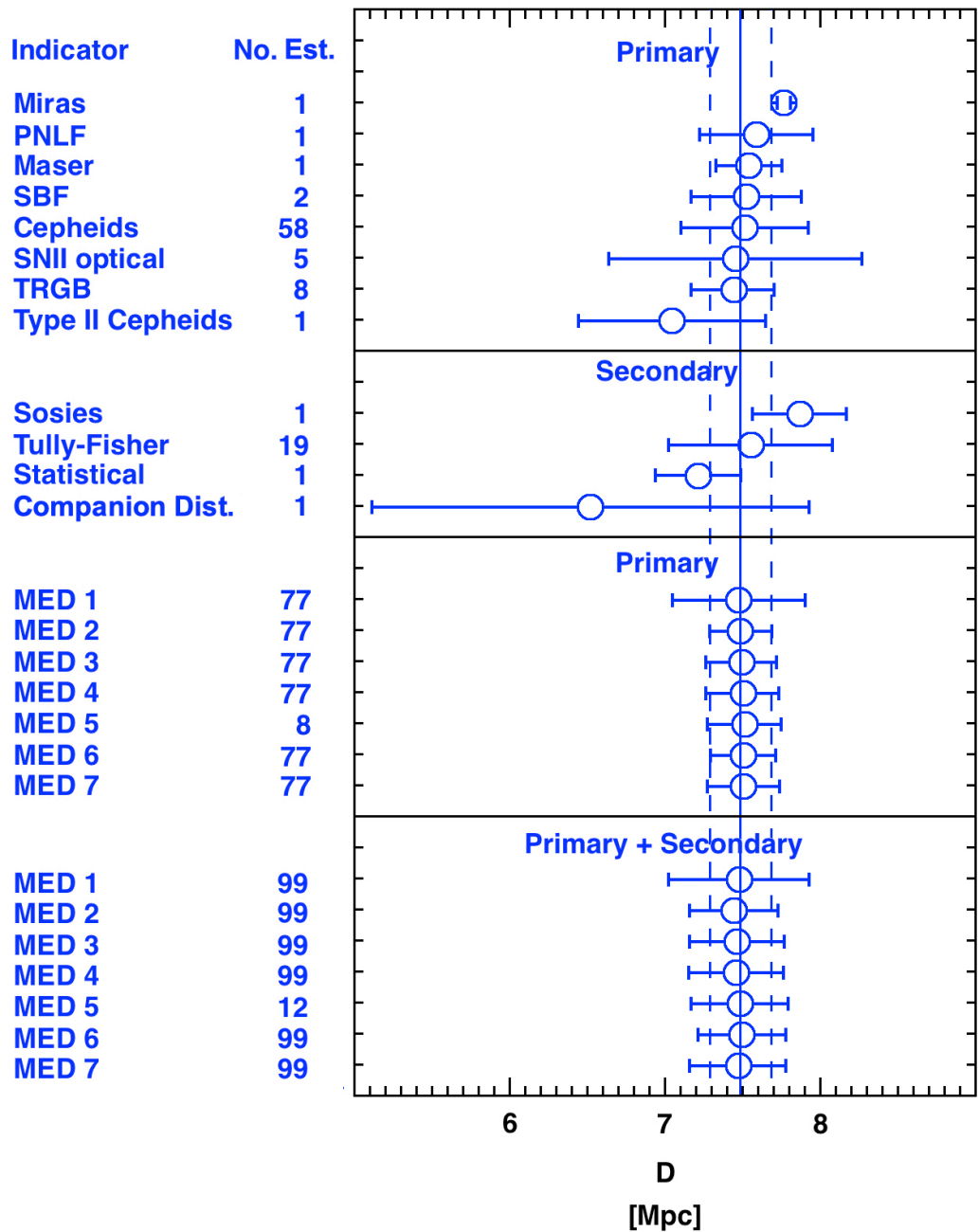

**Figure 3**. Messier 106 mean distance (solid line) and 1σ standard deviation (dashed lines) from MED method 2, the unweighted mean of eight primary indicators. Primary indicator means and standard deviations are shown by data points and error bars, except for indicators with one estimate, where 1σ precision of individual estimates is shown. Secondary indicators follow. Data points and error bars following show seven MED methods based on eight primary indicators, and based on eight primary and four secondary indicators, weighting primary over secondary at 4:1, as explained in text.

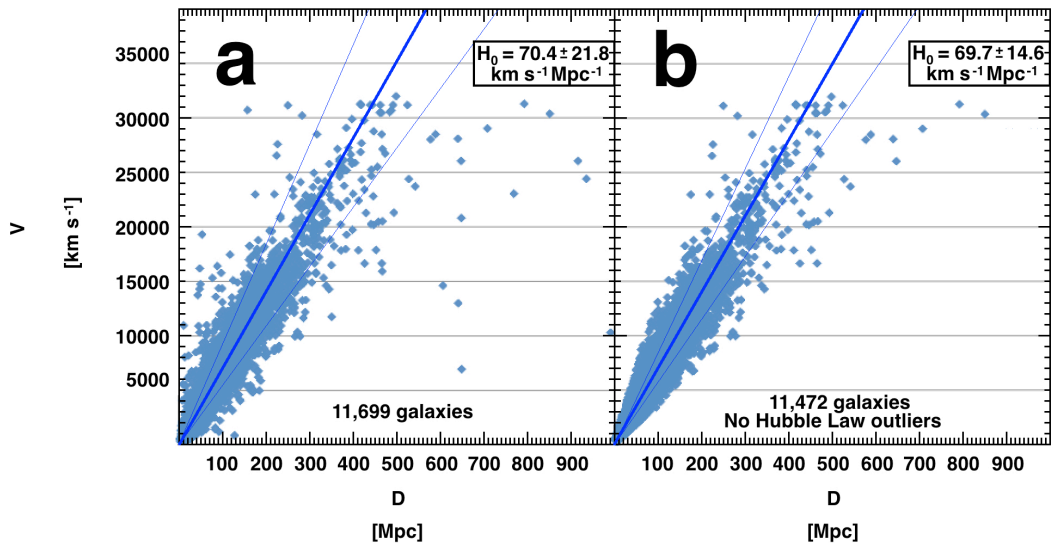

**Figure 4**. Hubble graph for 11,699 galaxies with multiple distances (a). The canonical standard deviation is sensitive to outliers, unlike say a median absolute deviation. Thus eliminating 2% of the Hubble Law outliers results in a sizable scatter reduction and yields $H_0 = 70 \pm 15$ km s$^{-1}$ Mpc$^{-1}$ (b).

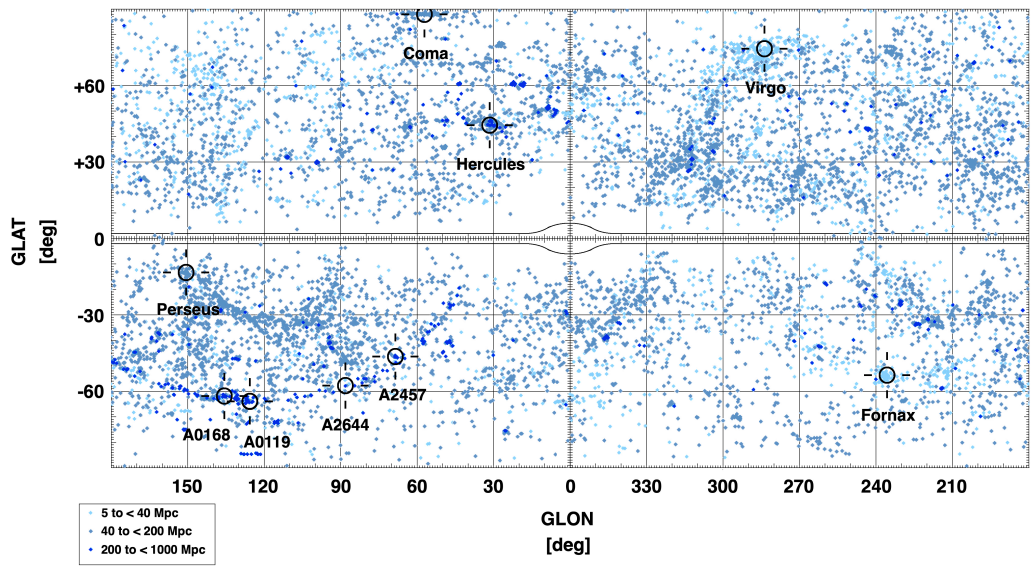

**Figure 5.** All-sky plot of 11,699 galaxies with multiple distances in galactic coordinates.

**Table 1**
Redshift-independent Distance Indicators (n = 77)

| Indicator | Tree Rank |
|---|---|
| **Primary Standard Candle** | |
| Asymptotic Giant Branch (AGB) Stars | 14 |
| Carbon Stars | 15 |
| Cepheids | 1 |
| Color-Magnitude Diagram (CMD) | 10 |
| Flux-Gravity-Luminosity Relation (FGLR) | 6 |
| Globular Cluster Luminosity Function (GCLF) | 9 |
| Horizontal Branch | 7 |
| Miras | 12 |
| Planetary Nebulae Luminosity Function (PNLF) | 13 |
| Red Clump | 4 |
| RR Lyrae | 3 |
| Surface Brightness Fluctuations (SBF) | 8 |
| Supernovae, Type 1a (SNIa) | 5 |
| Supernovae, Type 1a SDSS (SNIa SDSS) | 16 |
| Tip of the Red Giant Branch (TRGB) | 2 |
| Type II Cepheids | 11 |
| **Primary Standard Ruler** | |
| Eclipsing Binary | 1 |
| Globular Cluster (GC) Radius | 4 |
| Maser | 2 |
| Proper Motion | 5 |
| Supernovae, Type II (SNII Optical) | 3 |
| **Secondary** | |
| Active Galactic Nucleus (AGN) Time Lag | 13 |
| B Stars | 4 |
| Brightest Cluster Galaxy (BCG) | 12 |
| BL Lac Luminosity | 13.5 |
| Blue Supergiant | 4.5 |
| Brightest Stars | 5 |
| CO Ring Diameter | 14 |
| Companion Dist. | 3.5 |
| D-Sigma | 1.5 |
| Delta Scuti | 2 |
| Faber-Jackson | 2.5 |
| Fundamental Plane (FP) | 3 |
| Gravitational Lens (G Lens) | 16.5 |
| Globular Cluster (GC) SBF | 5.5 |
| Grav. Stability Gas. Disk | 14.5 |
| Gravitational Wave (Grav. Wave) | 19 |
| Gamma-Ray Burst (GRB) | 17 |
| Jet Proper Motion | 15.5 |
| M Stars | 6 |
| Magnitude | 18.5 |
| Novae | 6.5 |
| OB Stars | 7 |
| Post Asymptotic Giant Branch (PAGB) Stars | 7.5 |
| Quasar Spectrum | 15 |
| Red Supergiant Variable (RSV) Stars | 8 |
| Red Variable (RV) Stars | 8.5 |
| S Doradus Stars | 9 |
| Short Gamma-Ray Burst (SGRB) | 17.5 |
| Sosies | 12.5 |
| Statistical | 9.5 |
| Subdwarf Fitting | 10 |
| SX Phe Stars | 10.5 |
| Sunyaev–Zel'dovich (SZ) Effect | 16 |
| Tully-Fisher | 1 |
| White Dwarfs | 11 |
| Tully Est | 18 |
| Wolf-Rayet | 11.5 |
| **Tertiary** | |
| Black Hole | 0 |
| Diameter | 0 |
| Dwarf Ellipticals | 0 |
| Dwarf Galaxy Diameter | 0 |
| Globular Cluster (GC) FP | 0 |
| Globular Cluster (GC) K vs. (J-K) | 0 |
| GeV TeV Ratio | 0 |
| H I + Optical Distribution | 0 |
| H II Luminoisty Function (H II LF) | 0 |
| HI I region diameter | 0 |
| IRAS | 0 |
| L(Hβ)–σ | 0 |
| Low Surface Brightness (LSB) Galaxies | 0 |
| Mass Model | 0 |
| Orbital Mech. | 0 |
| Radio Brightness | 0 |
| Ring Diameter | 0 |
| Supernovae, Type II (SNII Radio) | 0 |
| Tertiary | 0 |

**Table 2**
Six Mean Estimate Distance (MED) Methods and a Combined Approach

| |
|---|
| MED 1 Unweighted distance estimates |
| MED 2 Unweighted distance indicators |
| MED 3 Error-weighted distance indicators |
| MED 4 Date-weighted distance indicators |
| MED 5 Selected distances per indicator |
| MED 6 Preference weighted distance indicators |
| MED 7 Weighted mean of MEDs 2, 3, 4, and 5 |

**Table 3**
Primary Indicator and Mean Estimate Distances for the LMC

| Indicator | No. Distances | Mean D (kpc) | Mean 1σ Std. Dev. (kpc) |
|---|---|---|---|
| AGB | 1 | 53.50 | 3.56 |
| Carbon Stars | 6 | 52.77 | 4.03 |
| Cepheids | 302 | 50.02 | 3.37 |
| CMD | 216 | 49.27 | 2.79 |
| FGLR | 1 | 50.10 | 3.58 |
| Horizontal Branch | 12 | 51.20 | 2.31 |
| Miras | 17 | 50.49 | 2.60 |
| PNLF | 10 | 48.71 | 2.43 |
| Red Clump | 74 | 48.28 | 2.57 |
| RR Lyrae | 173 | 49.34 | 2.93 |
| TRGB | 23 | 51.23 | 2.73 |
| Type II Cepheids | 22 | 50.35 | 2.14 |
| Eclipsing Binary | 41 | 48.55 | 2.49 |
| GC Radius | 1 | 50.10 | 2.60 |
| Maser | 3 | 47.03 | 2.14 |
| Proper Motion | 3 | 50.53 | 0.74 |
| SNII Optical | 35 | 49.85 | 2.87 |
| MED 1 | 940 | 49.57 | 3.03 |
| MED 2 | 940 | 50.08 | 1.58 |
| MED 3 | 940 | 50.29 | 1.37 |
| MED 4 | 940 | 50.31 | 1.45 |
| MED 5 | 17 | 49.90 | 1.78 |
| MED 6 | 940 | 49.33 | 1.58 |
| MED 7 | 940 | 50.09 | 1.61 |

**Table 4**
Primary and Secondary Indicator and Mean Estimate Distances for Messier 106

| Indicator | No. Distances | Mean D (Mpc) | Mean 1σ Std. Dev. (Mpc) |
|---|---|---|---|
| **Primary** | | | |
| Miras | 1 | 7.76 | 0.04 |
| PNLF | 1 | 7.59 | 0.36 |
| Maser | 1 | 7.54 | 0.21 |
| SBF | 2 | 7.52 | 0.35 |
| Cepheids | 58 | 7.51 | 0.41 |
| SNII optical | 5 | 7.45 | 0.81 |
| TRGB | 8 | 7.44 | 0.26 |
| Type II Cepheids | 1 | 7.05 | 0.60 |
| **Secondary** | | | |
| Sosies | 1 | 7.87 | 0.30 |
| Tully-Fisher | 19 | 7.55 | 0.53 |
| Statistical | 1 | 7.21 | 0.27 |
| Companion Dist. | 1 | 6.52 | 1.40 |
| **Primary** | | | |
| MED 1 | 77 | 7.47 | 0.42 |
| MED 2 | 77 | 7.48 | 0.20 |
| MED 3 | 77 | 7.49 | 0.23 |
| MED 4 | 77 | 7.50 | 0.23 |
| MED 5 | 8 | 7.52 | 0.23 |
| MED 6 | 77 | 7.50 | 0.20 |
| MED 7 | 77 | 7.51 | 0.23 |
| **Primary + Secondary** | | | |
| MED 1 | 99 | 7.48 | 0.44 |
| MED 2 | 99 | 7.45 | 0.28 |
| MED 3 | 99 | 7.46 | 0.30 |
| MED 4 | 99 | 7.46 | 0.30 |
| MED 5 | 12 | 7.48 | 0.31 |
| MED 6 | 99 | 7.50 | 0.28 |
| MED 7 | 99 | 7.47 | 0.30 |

**Table 5**
Mean Estimate Distances for 40 Messier Galaxies

| Source | NDist | MED2 | e_MED2 | MED3 | e_MED3 | MED4 | e_MED4 | MED5 | e_MED5 |
|---|---|---|---|---|---|---|---|---|---|
| MESSIER 031 | 411 | 0.756 | 0.05 | 0.757 | 0.052 | 0.758 | 0.054 | 0.76 | 0.062 |
| MESSIER 032 | 43 | 0.769 | 0.043 | 0.774 | 0.035 | 0.782 | 0.052 | 0.77 | 0.04 |
| MESSIER 033 | 165 | 0.853 | 0.065 | 0.856 | 0.065 | 0.856 | 0.065 | 0.856 | 0.075 |
| MESSIER 049 | 70 | 16.15 | 1.02 | 16.2 | 0.71 | 16.73 | 1.13 | 16.53 | 1.41 |
| MESSIER 051 | 53 | 8.03 | 0.9 | 7.96 | 0.86 | 7.98 | 0.86 | 7.74 | 1.07 |
| MESSIER 058 | 24 | 20.77 | 1.12 | 20.53 | 0.99 | 20.27 | 0.93 | 19.71 | 0.57 |
| MESSIER 059 | 47 | 15.72 | 1.08 | 15.65 | 0.91 | 15.67 | 1.04 | 15.4 | 0.91 |
| MESSIER 060 | 47 | 16.64 | 1.1 | 16.71 | 1.11 | 16.78 | 1.27 | 16.88 | 1.29 |
| MESSIER 061 | 14 | 16.34 | 4.13 | 16.68 | 4.27 | 15.6 | 3.94 | 17.83 | 1.32 |
| MESSIER 063 | 29 | 11.52 | 3.72 | 10.38 | 2.67 | 11.52 | 3.72 | 11.4 | 3.96 |
| MESSIER 064 | 25 | 5.99 | 1.14 | 5.5 | 1.03 | 5.51 | 0.59 | 5.21 | 0.56 |
| MESSIER 065 | 20 | 13.49 | 1.94 | 13.67 | 1.94 | 13.89 | 1.94 | 14.24 | 1.02 |
| MESSIER 066 | 86 | 10.08 | 0.55 | 9.68 | 0.8 | 10.3 | 0.75 | 9.3 | 1.29 |
| MESSIER 074 | 34 | 8.85 | 1.1 | 8.94 | 1.03 | 8.9 | 1.19 | 9.21 | 0.96 |
| MESSIER 077 | 13 | 17.3 | 3.71 | 17.22 | 3.82 | 17.2 | 3.85 | 17.2 | 3.84 |
| MESSIER 081 | 112 | 3.65 | 0.19 | 3.66 | 0.2 | 3.67 | 0.18 | 3.61 | 0.33 |
| MESSIER 082 | 21 | 3.7 | 0.15 | 3.64 | 0.19 | 3.62 | 0.17 | 3.6 | 0.24 |
| MESSIER 083 | 19 | 5.17 | 0.34 | 5.16 | 0.3 | 5.07 | 0.14 | 4.95 | 0.27 |
| MESSIER 084 | 62 | 17.12 | 0.81 | 17.12 | 0.89 | 17.08 | 0.93 | 17.49 | 0.99 |
| MESSIER 085 | 28 | 15.94 | 4.25 | 16.42 | 2.53 | 16.12 | 3.08 | 15.58 | 2.25 |
| MESSIER 086 | 54 | 16.27 | 1.25 | 16.44 | 1.19 | 16.47 | 1.68 | 16.76 | 1.38 |
| MESSIER 087 | 130 | 16.53 | 1.15 | 16.95 | 1.89 | 16.66 | 1.58 | 17.43 | 3.02 |
| MESSIER 088 | 43 | 15.25 | 2.84 | 15.95 | 2.73 | 15.56 | 3.01 | 15.87 | 0.38 |
| MESSIER 089 | 46 | 16.7 | 1.14 | 16.9 | 1.05 | 16.92 | 0.61 | 17.16 | 0.8 |
| MESSIER 090 | 14 | 16.61 | 1.87 | 16.41 | 2.16 | 16.44 | 2.11 | 16.39 | 2.19 |
| MESSIER 091 | 51 | 17.15 | 2.2 | 16.76 | 1.83 | 17.02 | 1.67 | 16.2 | 2.6 |
| MESSIER 094 | 25 | 4.95 | 0.44 | 4.94 | 0.38 | 4.89 | 0.36 | 4.97 | 0.42 |
| MESSIER 095 | 68 | 10.29 | 0.59 | 10.19 | 0.14 | 10.5 | 0.76 | 10.35 | 0.57 |
| MESSIER 096 | 74 | 10.55 | 0.85 | 10.42 | 0.92 | 10.52 | 1 | 10.21 | 0.91 |
| MESSIER 098 | 21 | 15.6 | 2.65 | 16.24 | 2.65 | 17.59 | 2.65 | 16.99 | 1.05 |
| MESSIER 099 | 15 | 14.94 | 1.96 | 14.86 | 1.84 | 14.65 | 2.14 | 14.67 | 2.11 |
| MESSIER 100 | 85 | 16.82 | 1.85 | 17.1 | 1.51 | 16.33 | 2.04 | 16.4 | 2.35 |
| MESSIER 101 | 134 | 6.9 | 0.5 | 6.97 | 0.6 | 6.86 | 0.55 | 6.95 | 0.68 |
| MESSIER 104 | 32 | 10.96 | 1.25 | 10.22 | 1.28 | 11.14 | 2.01 | 9.4 | 0.66 |
| MESSIER 105 | 60 | 10.47 | 0.91 | 10.62 | 0.76 | 10.66 | 0.85 | 10.89 | 0.55 |
| MESSIER 106 | 106 | 7.41 | 0.29 | 7.43 | 0.31 | 7.45 | 0.3 | 7.45 | 0.33 |
| MESSIER 108 | 23 | 13.19 | 2.71 | 13.18 | 2.71 | 12.82 | 3.22 | 13.75 | 1.91 |

| | | | | | | | | |
|---|---|---|---|---|---|---|---|---|
| MESSIER 109 | 26 | 20.83 | 1.77 | 20.83 | 1.77 | 21.72 | 1.77 | 20.58 | 2.11 |
| MESSIER 110 | 41 | 0.814 | 0.042 | 0.827 | 0.044 | 0.812 | 0.038 | 0.825 | 0.039 |
| MESSIER 102[a] | 13 | 14.23 | 1.23 | 14.33 | 1.14 | 14.13 | 1.43 | 14.41 | 1.03 |
| Total/Mean[b] | 2378 | 11.63 | 1.37 | 11.63 | 1.28 | 11.69 | 1.39 | 11.65 | 1.19 |

**Notes.** This table is available in its entirety in machine-readable form.

[a] Messier 102 ID as NGC 5866 to be confirmed.
[b] The last row, Source = "Total/Mean," includes the total number of observations and the mean of 40 values listed in each column of the table.

**Table 6**
Mean Estimate Distances for 11,699 Galaxies

| Seq | Source | GLON | GLAT | z | cz | BED1 | e_BED1 | BED2 | e_BED2 | BED3 | e_BED3 | BED4 | e_BED4 | BED5 | e_BED5 | BED6 | e_BED6 | BED7 | e_BED7 | Nest | $H_0$ | $v_{pec}$ |
|---|---|---|---|---|---|---|---|---|---|---|---|---|---|---|---|---|---|---|---|---|---|---|
| 1 | NGC 5332 | 0.159 | 72.678 | 0.022416 | 6645 | 103.3 | 17.7 | 98.2 | 15.3 | 97.5 | 14.6 | 98.2 | 15.3 | 92.0 | 9.1 | 110.4 | 15.3 | 95.3 | 12.2 | 3 | 69.7 | 29 |
| 2 | NGC 5839 | 0.200 | 49.009 | 0.004069 | 1217 | 23.8 | 0.8 | 23.8 | 0.2 | 24.2 | 0.2 | 24.3 | 0.2 | 24.4 | 0.2 | 24.0 | 0.2 | 24.3 | 0.3 | 6 | 50.0 | 486 |
| 3 | CGCG 075-013 | 0.239 | 63.845 | 0.025818 | 7640 | 134.3 | 3.3 | 134.3 | 3.3 | 133.7 | 3.3 | 133.1 | 3.3 | 133.0 | 3.3 | 134.3 | 3.3 | 133.2 | 3.3 | 4 | 57.3 | 1686 |
| 4 | ESO 346- G 014 | 0.272 | -63.174 | 0.008980 | 2680 | 46.6 | 6.4 | 46.6 | 6.4 | 48.8 | 6.4 | 46.3 | 6.4 | 48.7 | 6.4 | 46.6 | 6.4 | 47.9 | 6.4 | 18 | 56.0 | 672 |
| 5 | UGC 09199 | 0.326 | 63.358 | 0.025808 | 7637 | 107.0 | 10.0 | 107.0 | 10.0 | 106.1 | 10.0 | 104.4 | 10.0 | 105.3 | 10.0 | 107.0 | 10.0 | 105.3 | 10.0 | 6 | 72.5 | -267 |
| 6 | NGC 6849 | 0.329 | -30.818 | 0.020147 | 5979 | 59.9 | 4.8 | 59.9 | 4.8 | 60.0 | 4.8 | 58.9 | 4.8 | 58.5 | 4.8 | 59.9 | 4.8 | 58.9 | 4.8 | 4 | 101.5 | -1854 |
| 7 | NGC 5845 | 0.339 | 48.904 | 0.004910 | 1468 | 28.0 | 2.7 | 28.0 | 0.1 | 29.4 | 0.1 | 30.7 | 0.1 | 31.5 | 0.1 | 28.1 | 0.1 | 30.7 | 0.3 | 5 | 47.8 | 682 |
| 8 | NGC 5846 | 0.426 | 48.797 | 0.005711 | 1707 | 27.5 | 3.1 | 27.7 | 2.3 | 27.9 | 2.2 | 27.0 | 1.4 | 28.2 | 2.4 | 27.9 | 2.3 | 27.8 | 2.2 | 23 | 61.5 | 236 |
| 9 | NGC 5681 | 0.452 | 58.970 | 0.026348 | 7795 | 116.6 | 9.5 | 116.6 | 9.5 | 118.4 | 9.5 | 111.3 | 9.5 | 115.0 | 9.5 | 116.6 | 9.5 | 114.5 | 9.5 | 3 | 68.1 | 223 |
| 10 | NGC 5850 | 0.516 | 48.636 | 0.008526 | 2545 | 25.4 | 4.0 | 25.4 | 4.0 | 23.6 | 4.0 | 18.8 | 4.0 | 17.8 | 4.0 | 25.4 | 4.0 | 19.5 | 4.0 | 6 | 130.8 | -1184 |

**Note.** This table is published in its entirety in the electronic version of this paper and is available in a machine-readable form. A portion is provided here to guide in its form and content.